\documentstyle[12pt]{article}

\newcommand{\sss}{\vspace{.2in}}
\begin{document}
{}~\hfill{UICHEP-TH/95-2}

\vspace{.3in}
\centerline{\large\bf Negaton and Positon Solutions of the KdV Equation}
\vspace{.6in}

\centerline{\bf C. Rasinariu, U. Sukhatme}
\centerline{\bf Department of Physics (m/c 273)}
\centerline{\bf University of Illinois at Chicago}
\centerline{\bf 845 W. Taylor Street}
\centerline{\bf Chicago, Illinois 60607-7059, U.S.A.}
\vspace{.3in}
\centerline{\bf Avinash Khare}
\centerline{\bf Institute of Physics}
\centerline{\bf Sachivalaya Marg}
\centerline{\bf Bhubaneswar 751005, India}
\vspace{.5in}

{\bf Abstract}
\sss

We give a systematic classification and a detailed discussion of the
structure, motion and scattering of the recently discovered
negaton and positon solutions of the
Korteweg-de Vries equation. There are two distinct types of negaton
solutions which we label $[S^{n}]$ and $[C^{n}]$, where $(n+1)$ is
the order
of the Wronskian used in the derivation. For negatons, the number
of singularities and zeros is finite and they show very interesting
time dependence. The general motion is in the positive $x$ direction,
except for certain negatons which exhibit one oscillation around the
origin.
In contrast, there is just one type of positon
solution, which we label $[\tilde C^n]$.
For positons, one gets a finite number of singularities
for $n$ odd, but an infinite number for even values of $n$. The general
motion of positons is in the negative $x$ direction with periodic
oscillations. Negatons and positons retain their identities in a
scattering process and their phase shifts are discussed. We obtain a
simple explanation of all phase shifts by generalizing the notions of
``mass" and ``center of mass" to singular solutions.
Finally, it is shown that negaton and positon
solutions of the KdV equation can be
used to obtain corresponding new solutions of the modified KdV equation.

\thispagestyle{empty}

\newpage

\centerline{\bf 1. Introduction}\sss

One of the most studied
nonlinear evolution equations in mathematical physics is
the Korteweg-de Vries (KdV) equation
\begin{equation}\label{kdv}
u_t-6uu_x+u_{xxx}=0.
\end{equation}
It is well-known that the KdV equation is completely integrable and
gives rise to  an infinite number of conservation laws
\cite{Drazin,Lamb,Eckhaus}. Although a great deal is
known about non-singular
multi-soliton solutions, singular solutions of the KdV equation
have been discussed to a much lesser
extent\cite{Kruskal74,Matveev92,Matveev94,Stahlhofen95}.
To our knowledge, a comprehensive
treatment of singular solutions
is not available. All the above solutions as well as more
complicated new solutions called negatons and positons
can all be obtained from Matveev's recent
generalized Wronskian formula for solutions of the KdV equation
\cite{Matveev92}. This formula makes use of
an arbitrary number of solutions
of the Schr\"odinger equation at energies $k_i^2$ and their derivatives
with respect to $k_i$ as inputs. In this paper, we only consider the
simplest case of a zero background potential, that is the free
particle Schr\"odinger equation.
If only one input solution at energy $k^2$ is used and it
has negative (positive) energy, the resulting KdV
solutions are called negatons (positons). Using several input solutions
permits the study of scattering. Similar approaches can also be
applied to other nonlinear
evolution equations \cite{Stahlhofen92,Beutler93,Beutler94}.

In this paper, we make a systematic
classification of negatons and positons for the KdV equation
and study their structure, motion
and interactions. We develop a physical picture underlying negaton and
positon solutions which helps to give an intuitive understanding of
their $x$ and $t$ dependences. Similarly, we generalize
the standard concepts of
``mass" and ``center of mass" to non-singular
solutions, and use them to give simple quantitative
explanations for the phase shifts in various scattering processes. We give
many figures showing negatons and positons in motion, since this
provides a good pictorial grasp of time dependence.
In Sec. 2, we present the general formalism and establish
notation relevant to solutions of the KdV equation. Sec. 3 contains a
detailed description and classification of negaton solutions. There are
two types of negatons $[C^{n}]$ and $[S^{n}]$, $n=0,1,2,\ldots$. The
singularity patterns and their time dependence are particularly interesting
and we describe these in detail. Sec. 4 contains a description of the
structure and motion of positons. There is only
one type of positon $[\tilde C^n]$, $n=0,1,2,\ldots$ and one has a finite
number of singularities for odd values of $n$. The motion for this
case is physically quite different from the negaton case.
This can be understood, since one is using trigonometric functions instead
of hyperbolic functions. Scattering of negatons and positons is treated in
Sec. 5. It is found that negatons and positons emerge from
an interaction preserving their identity, but often with a shift in
phase. Finally, in Sec. 6, we discuss the positon and negaton solutions
of the modified Korteweg-de Vries (mKdV) equation. Some open
problems and concluding remarks are given in Sec. 7. \sss

\centerline{\bf 2. General Formalism and Notation}\sss

Solutions of Eq. (\ref{kdv}) can be systematically obtained from solutions
of the free particle Schr\"odinger equation ($\hbar=2m=1$):
\begin{equation}\label{schr}
-\frac{d^2\phi_i}{dx^2}=E_i\phi_i.
\end{equation}
For $E_i=-k_i^2<0$, a convenient choice of
independent solutions $\phi_i(x)$ is $\sinh k_ix$ and
$\cosh k_ix$, whereas for
$E_i=\tilde k_i^2>0$, the corresponding trigonometric functions
$\sin \tilde k_ix$ and $\cos \tilde k_ix$ can be chosen.
(It can be shown that nothing new is obtained by taking more
general linear combinations).
We consider solutions of Eq. (\ref{kdv})
of the form \cite{Matveev92,Matveev91}
\begin{equation}\label{uxt}
u(x,t)=-2\frac{\partial ^2}{\partial x^2} \ln W =
2 \frac{({W'}^2-WW'')}{W^2}~~,
\end{equation}
where $W=W(\phi_1,\ldots,\phi_n)$ is the Wronskian determinant composed of
$\phi_i(\theta_i)$. Here $\theta_i$ stands for:
\begin{equation}\label{thetan}
\theta_i=k_i(x+\xi_i(k_i)-4k^2_it)
\end{equation}
for $E_i<0$, and
\begin{equation}\label{thetap}
\tilde \theta_i=\tilde k_i(x+\tilde \xi_i(\tilde k_i)+4\tilde k^2_it)
\end{equation}
for $E_i>0$. $\xi_i(\tilde \xi_i)$ are arbitrary functions of
$k_i(\tilde k_i),~~i=1,2,\ldots,n$.

For clarity, let us first focus on solutions of the KdV equation
which come from negative energy solutions of Eq. (\ref{schr}). The
simplest choice is to have a Wronskian of order 1. Here, we have
two types of solutions which can be either $\cosh
\theta$ or $\sinh \theta$. For $\phi=\cosh \theta$, one gets $u(x,t)=-2k^2
{\rm sech}^2 \theta$,  which we denote by $[C^0]$.
This is the usual nonsingular one
soliton solution moving to the right  (along the positive $x$ direction)
with speed $4k^2$. For $\phi=\sinh \theta$, one
gets $u(x,t)=2k^2 {\rm cosech}^2 \theta$, which we
denote by $[S^0]$. This is the simplest singular solution of the KdV
equation \cite{Drazin}. There is just one singularity at $x=4k^2t-\xi(k)$
moving to the right at speed $4k^2$. $[C^0]$ and $[S^0]$ are called negaton
solutions of order 0.

For Wronskians of order 2 there are three types of solutions:
$${\rm (a)}~~~\phi_1=\cosh \theta_1,~~~\phi_2=\cosh \theta_2~~;$$
$${\rm (b)}~~~\phi_1=\cosh \theta_1,~~~\phi_2=\sinh \theta_2~~;$$
$${\rm (c)}~~~\phi_1=\sinh \theta_1,~~~\phi_2=\sinh \theta_2~~;$$
where $\theta_1,\theta_2$ are given by Eq. (\ref{thetan}), and
correspond to speeds $4k_1^2$ and $4k_2^2$
respectively. It is easy to
check \cite{Matveev92} that case (b) is the well-known finite
two soliton solution of the KdV
equation,  whereas cases
(a) and (c) correspond to solutions with one singularity.

Of particular interest for us is the situation where $k_1=k$ and $k_2=k+
\epsilon$ with $\epsilon$ tending to zero. In order to get a non-trivial
solution, it is necessary to make the choice $\xi_1(k)=\xi_2(k)=\xi(k)$
in Eq. (\ref{thetan}).
For cases (a) and (c), $W$, $W'$ and $W''$ are all
$O(\epsilon)$. Thus, from Eq. (\ref{uxt}),
$u(x,t)$ is $O(\epsilon^0)$. This is a new solution \cite {Matveev92}
of the KdV equation which
does not vanish as $\epsilon \rightarrow 0$. For case (b), however,
$W \rightarrow$ constant as $\epsilon \rightarrow 0$ and no new solutions
result.

The new solutions coming from case (a) and case (c)
will be denoted by $[C]$
and $[S]$ respectively and are called negaton solutions of order 1.
More explicitly, for these
cases $\phi_2(k+\epsilon)
=\phi_1(k)+\epsilon \partial_k\phi_1(k)+O(\epsilon^2)$,
and the Wronskian is
\begin{equation}\label{W2}
W(\phi_1,\phi_2) \equiv W(\phi_1, \phi_1+\epsilon \partial_k\phi_1)
= \epsilon W(\phi_1, \partial_k \phi_1).
\end{equation}
The multiplicative constant $\epsilon$ does not play any role in
obtaining $u(x,t)$ using Eq. (\ref{uxt}) and can be dropped
from the Wronskian. For the $[C]$ case,
\begin{equation}\label{wc}
W \equiv W(\cosh \theta, \partial_k \cosh \theta)
=k\gamma+\cosh\theta \sinh \theta,
\end{equation}
where
\begin{equation}\label{gamma}
\gamma \equiv \partial_k\theta=x+\xi(k)+k \partial_k\xi(k)-12k^2t .
\end{equation}
Similarly, for the $[S]$ case, the Wronskian reads:
\begin{equation}\label{ws}
W \equiv W(\sinh \theta, \partial_k \sinh \theta)
=-k\gamma+\cosh\theta \sinh \theta.
\end{equation}

Although we have so far only discussed negaton Wronskians of order 1 and 2,
the above results can be readily extended to Wronskians of any higher
order. A straightforward extension of
Eq. (\ref{W2}) yields a Wronskian determinant of order $(n+1)$:
\begin{equation}\label{Wn}
W=W(\phi, \partial_k \phi, \ldots,\partial_k^{n} \phi).
\end{equation}
This is a special case of the generalised Wronskian formula given by
Matveev \cite{Matveev92}. If the Wronskian of Eq. (\ref{Wn}) with
$\phi=\cosh \theta$ is used in Eq. (\ref{uxt}), the resulting KdV
solution is called a negaton $[C^n]$ of order $n$ with $n=0,1,2,...$.
Similarly, the choice $\phi=\sinh \theta$ yields a
negaton $[S^n]$ of order $n$.
To summarize, the negaton corresponding to
Eq. (\ref{Wn}) has the physical interpretation of merging $(n+1)$ solutions
$\phi$ of the free particle Schr\"odinger equation all with wave numbers
near $k$ and identical phases $\xi (k)$.

The entire discussion given above for negatons
also holds for positive energy solutions
of the Schr\"odinger equation. The solutions of the KdV
equation resulting from the choices $\phi=\cos \tilde \theta$  and
 $\phi=\sin \tilde \theta$
are called positons of order $n$ \cite{Matveev92} and are denoted by
$[\tilde C^n]$ and
$[\tilde S^n]$ respectively.

An important difference between positons and negatons is
that the positons $[\tilde C^n]$ and  $[\tilde S^n]$ are not independent. In
fact, the choice $\tilde k\xi_i=\pi/2$ in $\tilde \theta_i$
in Eq. (\ref{thetap})
transforms  $[\tilde C^n]$ into  $[\tilde S^n]$. On the other hand, negatons
 $[C^{n}]$ and  $[S^{n}]$ are physically different. As we shall see, they
usually have a different number of singularities for the same
value of $n$. However, one can mathematically transform  $[C^{n}]$
into  $[S^{n}]$ by the unphysical imaginary choice of phase $k\xi_i=i \pi/2$.

It is also interesting to observe that
positon solutions can be obtained from the corresponding negaton
solutions via the change $k \rightarrow i\tilde k$.
Note that the $x$ and $t$ dependence of all solutions
comes from $\theta~(\tilde \theta)$
and derivatives of $\theta~(\tilde \theta)$ with respect to $k~(\tilde k)$.
 From now on, our discussion is based on making the simplest
choice $\xi(k)=0$ in Eqs.
(\ref{thetan}), (\ref{thetap}) and (\ref{gamma}).
It is important to observe that under the transformations
$x \rightarrow -x$ and $t \rightarrow -t$, $\theta~(\tilde \theta)$
and all derivatives with respect to $k~(\tilde k)$ change sign. As a
result, the Wronskian $W$ in Eq. (\ref{Wn}) has the
property $W(-x,-t)=\pm W(x,t).$
Thus, for all negaton or positon solutions, it follows that
$u(x,t)=u(-x,-t)$, and it
is sufficient to just consider the behavior at either negative or
positive values of $t$. In particular, at time $t=0$, all solutions
are symmetric $u(x,0)=u(-x,0)$.

The ``mass" and ``center of mass" of any solution $u(x,t)$
of the KdV equation are useful concepts in analyzing the behavior
of negatons and positons. Here, $u(x,t)$ is identified with a
linear mass distribution, and the total mass is given by
\begin{equation}
M\equiv\int^\infty_{-\infty}u(x,t)dx.
\end{equation}
This definition is only useful for nonsingular solutions $u(x,t)$. However,
it is easy to obtain an alternative, more generally applicable definition.
Using Eq. (\ref{uxt}) for nonsingular $u(x,t)$, the total mass can be
written as
\begin{equation}\label{M}
M=-2[W'/W]^{+\infty}_{-\infty}~~.
\end{equation}
We will use Eq. (\ref{M}) as the definition of the mass $M$,
an expression which is well-defined for both nonsingular as well as
singular solutions $u(x,t)$. $M$ is a constant of the motion.
Note that our definition is equivalent to
the $x+i\epsilon$ regularization procedure suggested
in Ref. \cite{Stahlhofen95}. Also , the position
of the center of mass is given by
\begin{equation}\label{xcm}
x_{CM}\equiv\frac{1}{M} \int^\infty_{-\infty} xu(x,t)dx.
\end{equation}
Again, using Eq. (\ref{uxt}), it is possible to
re-write the expression for
the center of mass position,
\begin{equation}
x_{CM}=
\frac{1}{M} \bigg [ -2x\frac{W'}{W}+ \ln W^2 \bigg ]^{+\infty}_{-\infty}.
\end{equation}
This definition can be used for all solutions $u(x,t)$. The center of mass
moves at a constant speed \cite{Drazin}.\sss

\centerline{\bf 3. Structure and Motion of Negatons}\sss

In this section, we describe the $x$ and $t$ dependences of
negatons $u(x,t)$
corresponding to Wronskians of different orders. A summary of some
characteristics and properties of the simplest
negatons is given in Table 1. \sss

\noindent {\bf Wronskians of order 1:} Here, one has the familiar results:
\begin{equation}
[C^0] ~~~~~u(x,t)=-2k^2{\rm sech}^2 \theta,
\end{equation}
\begin{equation}
[S^0] ~~~~~u(x,t)=2k^2{\rm cosech}^2 \theta.
\end{equation}
Both negatons move with constant speed $4k^2$,
and their shape remains unchanged. The ``masses" of both the $[C^0]$ and
$[S^0]$ negatons as given by Eq. (\ref {M}) are $-4k$. \sss

\noindent {\bf Wronskians of order 2:} The explicit Wronskians are given
in Table 1 and the corresponding KdV solutions exhibit very
interesting behavior. The $[C]$
negaton is given by:
\begin{equation}\label{ucc}
[C]~~~~~u(x,t)=\frac{8k^2\cosh \theta(\cosh \theta-k
\gamma \sinh \theta)} {(\cosh\theta \sinh\theta+k \gamma)^2}.
\end{equation}
Its shape and motion is shown in Fig. 1.
It has one singularity corresponding to the zero of its Wronskian (see
Table 1). At any fixed time $t$, the dominant term in the
Wronskian at $x \rightarrow \pm \infty$ is
$\cosh\theta \sinh\theta$. Therefore, one expects the Wronskian to
necessarily have
an odd number of zeros. For this case, there is just
one zero giving rise to the singular behavior
$u(x,t)\propto \frac{2}{(x-x_P(t))^2}$.
At large negative time $t$, since the main term in the Wronskian is
$\cosh\theta\sinh\theta$, one gets a $[C]$ negaton composed of a
``soliton" $[C^0]$ (corresponding to the $\cosh\theta$ factor) with a
singularity $[S^0]$ on the left (corresponding to the $\sinh\theta$ factor).
This structure immediately suggests that
the mass of the $[C]$ negaton should be $-8k$, and a computation using
Eq. (\ref{M}) confirms this to be the case.
The ``center of mass" of the negaton is approximately half way
between the singularity and the minimum of the ``soliton", and it
moves with a constant speed $4k^2$.
The motion of the pole $x_P(t)$ is shown in Fig. 2, which shows its position
and speed. Note
that the pole has an asymptotic speed $4k^2$, which is expected since the
Wronskian just becomes a function of $\theta$ at $t \rightarrow \pm \infty$.
$u(x,t)$ also has two zeros coming from the numerator of
Eq. (\ref{ucc}). These two zeros move as shown in Fig. 3.

Similarly, the $[S]$ negaton is
\begin{equation}\label{uss}
[S]~~~~~u(x,t)=\frac{-8k^2\sinh \theta(\sinh \theta-k
\gamma \cosh \theta)} {(\sinh\theta \cosh\theta-k \gamma)^2}.
\end{equation}
This is similar in form to the $[C]$ negaton with $\sinh\theta$ and
$\cosh\theta$ exchanged. As can be seen in Fig. 4, at large negative time, the
$[S]$ negaton is a singularity $[S^0]$  (corresponding to
the $\sinh\theta$ factor in $W$) along with a ``soliton" $[C^0]$ on the
left  (corresponding to the $\cosh\theta$ factor in $W$). It has a
mass $-8k$.
The $[S]$ negaton shows
an interesting oscillation of its singularity near $x=t=0$.
The singularity
moves continuously to the right and comes to a momentary
halt at a positive value of $x$.
It then reverses its direction of motion, goes past the
origin at time $t=0$ with infinite instantaneous velocity
and again comes to a halt at a negative value
of $x$.  Thereafter, the motion of the singularity is continuously in
the positive $x$ direction, with an asymptotic speed $4k^2$.
The motion of the $[S]$
negaton and the time-dependence of its singularity and its two zeros
are shown in Figs. 4, 5 and 6. \sss

\noindent {\bf Wronskians of order 3,4 and 5:} The Wronskians
of order 3 and 4 are given by:
\begin{equation}\label{wc3}
[C^2]~~~~~W={1 \over 2}\sinh {3\theta}+\sinh{\theta}({1 \over 2} +4k^2
\gamma^2)-2k\gamma\cosh{\theta}+48k^3t\cosh{\theta},
\end{equation}
\begin{equation}\label{ws3}
[S^2]~~~~~W={1 \over 2}\cosh {3\theta}-\cosh{\theta}({1 \over 2}+4k^2
\gamma^2)+2k\gamma\sinh{\theta}-48k^3t\sinh{\theta},
\end{equation}
\[
[C^3]~W={3 \over 2}\cosh {4\theta}+\cosh{2\theta}(-24k^2
\gamma^2+576k^4\gamma t)
+\sinh{2\theta}(16k^3\gamma^3+12k\gamma-384k^3t)
\]
\begin{equation}\label{wc4}
-16k^4\gamma^4-12k^2\gamma^2-192k^4\gamma t-6912k^6t^2-{3 \over 2},
\end{equation}
\[
[S^3]~W={3 \over 2}\cosh {4\theta}+\cosh{2\theta}(24k^2
\gamma^2-576k^4\gamma t)
-\sinh{2\theta}(16k^3\gamma^3+12k\gamma-384k^3t)
\]
\begin{equation}\label{ws4}
-16k^4\gamma^4-12k^2\gamma^2-192k^4\gamma t-6912k^6t^2-{3 \over 2}.
\end{equation}
For any Wronskian, the corresponding KdV solution is readily obtained
from Eq. (\ref{uxt}). The number of zeros and singularities is shown in
Table 1. In particular, the $[S^4]$ negaton has eight zeros
and three singularities. Their motion is shown in Figs. 7 and 8.
Again, note that at large negative time, the $[S^4]$ negaton has
a dominant term in the Wronskian
``$\sinh\theta\cosh\theta\sinh\theta\cosh\theta\sinh\theta$",
which gives the structure
``singularity-soliton-singularity-soliton-singularity" seen in Fig. 7.
The two leading singularities show an oscillation around $x=t=0$, but
the third one does not.
\sss

\noindent {\bf Wronskians of order {\bf $(n+1)$}:} At this stage, let us
generalize the discussion to Wronskian determinants of
arbitrary order $(n+1), n=0,1,2,...$. For any given negaton, the
number of singularities and the number of zeros are both finite.
These numbers become steadily larger as the
order of the Wronskian
increases. The number of singularities and the number of
zeros remains constant in
time and hence characterize a given negaton.
The dominant terms in the Wronskians at
$x \rightarrow \pm \infty$ are given in Table 1. They follow a simple
rule, which tells whether there are an odd or an even number of negaton
singularities. Based on these considerations, we expect the following
general formulas for the number of singularities:
$$
[C^{n}]~~~ (n+1)/2~~{\rm for}~n~ {\rm odd}~;~~ n/2~~{\rm for} ~n~ {\rm even}~;
$$
\begin{equation}\label{poles}
[S^{n}]~~~ (n+1)/2~~{\rm for}~n~ {\rm odd}~;~~ (n+2)/2~~{\rm for}
{}~n~ {\rm even}~.
\end{equation}
Similar considerations show that there are $2n$ zeros for both $[C^{n}]$
and $[S^{n}]$ negatons. At least for the choice $\xi(k)=0$, we have checked
that the
number of zeros remains unchanged in time. It is easy to show that the mass
of the $[C^{n}]$ and $[S^{n}]$ negatons is $-4(n+1)k$, and the
center of mass moves with constant speed $4k^2$.

It is interesting to analyze some features of negatons at time $t=0$.
The Wronskian for $[S^{n}]$ has the flat behavior $(kx)^{(n+1)(n+2)/2}$
 near $x=0$ and consequently the KdV solution has the behavior
$u(x,0)~\propto~\frac{(n+1)(n+2)}{x^2}$. Likewise, $[C^{n}]$ has the
singular behavior $u(x,0)~\propto~\frac{n(n+1)}{x^2}$ at small $x$.
Note that the time $t=0$ is very special, since all negaton
singularities merge at $x=0$. At any other time $t$, the singularities
are separated, each exhibiting a $\frac{2}{(x-x_P(t))^2}$ behavior.
For any given negaton, the separation between singularities
becomes constant at large $t$, since all singularities
asymptotically move at the same speed $4k^2$.\sss

\centerline{\bf 4. Structure and Motion of Positons}\sss

The analysis for positons is somewhat different than for negatons since
there is only one type labeled by $[\tilde C^n]$ and trigonometric
functions are involved. A summary of properties is
given in Table 2. The simplest positon is
\begin{equation}
[\tilde C^0]~~~ W=\cos \tilde \theta~,~~~u(x,t)
=2\tilde k^2\sec^2 \tilde \theta~~,
{}~~\tilde \theta =\tilde k(x+4\tilde k^2t),
\end{equation}
which moves to the left at a constant speed $4\tilde k^2$. It
has an infinite
number of singularities, and in fact this property holds for positons
of any even order $n$ \cite{Stahlhofen95}. In contrast, for odd values
of $n$, the number of singularities is finite but the number of zeros
is infinite. The positon of order 1
$[\tilde C]$ has been extensively discussed \cite {Matveev92,Stahlhofen95}.
\begin{equation}\label{utcc}
[\tilde C]~~~~~u(x,t)=\frac{8\tilde k^2\cos \tilde \theta(\cos \tilde \theta
+\tilde k\tilde \gamma \sin \tilde \theta)}
{(\sin \tilde \theta \cos \tilde \theta+\tilde k \tilde \gamma)^2},
\end{equation}
where $\tilde \gamma \equiv \partial_{\tilde k} \tilde \theta$.
In order to compare the motion of positons with negatons,
we show the motion of the singularity of the $[\tilde C]$ positon
in Fig. 9. The graph shows several more or less straight sections with
periodic jumps. The straight sections have an average
slope $8\tilde k^2$ corresponding to the difference of the two characteristic
speeds $4\tilde k^2$ and $12\tilde k^2$ contained in
the quantities $\tilde \theta$ and
$\tilde \gamma $ respectively in the Wronskian. The jumps occur at times
$(2m+1) \pi /(16\tilde k^3)$ for $m=0, \pm 1, \ldots$,
and give rise to infinite speeds. Alternatively, the motion of the
singularity can also be described as oscillations around an average
constant speed $12 \tilde k^2$.
A more detailed description of this motion
and an extension to other even values of $n$ can be found
in Ref. \cite{Stahlhofen95}.
For completeness, we give expressions
for the Wronskians for $[\tilde C^2]$ and $[\tilde C^3]$ positons:
\begin{equation}\label{wtc3}
[\tilde C^2]~~~~~W={1 \over 2}\sin {3\tilde \theta}+\sin{\tilde \theta}
({1 \over 2} -4\tilde k^2
\tilde \gamma^2)-2\tilde k \tilde \gamma\cos{\tilde \theta}
-48\tilde k^3t\cos{\tilde \theta}~~,
\end{equation}
\[
[\tilde C^3]~~~~~W=\cos{2\tilde \theta}(-24\tilde k^2
\tilde \gamma^2-576\tilde k^4 \tilde \gamma t)
+\sin{2\tilde \theta}(-16\tilde k^3 \tilde \gamma^3+12\tilde k
\tilde \gamma+384\tilde k^3t)
\]
\begin{equation}\label{wtc4}
-{3 \over 2}\cos{4\tilde \theta}+16\tilde k^4\tilde \gamma^4
-12\tilde k^2 \tilde \gamma^2+192\tilde k^4 \tilde \gamma t
-6912\tilde k^6t^2+{3 \over 2}~~.
\end{equation}
The mass of any positon with odd $n$ is zero. This follows
from Eq. (\ref{M}) since all Wronskians of odd order $n$ have a powerlike
behavior for $x \rightarrow \pm \infty$.
\sss

\centerline{\bf 5. Scattering of Negatons and Positons}\sss

Now that we have classified the various types of negaton and positon solutions
of the KdV equation and studied their individual structures and motions, we
proceed to a discussion of scattering.
For simplicity, we restrict our attention to
processes involving two incident objects (negatons or positons)
with wave numbers $k_1$ and $k_2>k_1$. As might be expected from
previous work, all these objects emerge from the scattering process
preserving their identity but often acquiring a phase
shift \cite{Drazin,Matveev92}. \sss

{\bf Negaton-negaton scattering:}  The simplest situation is the scattering
of two negatons of order 0. There are four possibilities which are shown in
Table 3. Contained therein is the standard nonsingular soliton-soliton
case $[C^0][C^0]$ resulting from the Wronskian
$W(\phi_1,\chi_2)$, where
\begin{equation}\label{phi}
\phi_i \equiv \cosh \theta_i~~,~\chi_i \equiv \sinh \theta_i~~,~ \theta_i=
k_ix-4k_i^3t~~,~i=1,2,\ldots.
\end{equation}
In general, for $N$ solitons, the asymptotic solution
at $t \rightarrow \pm \infty$ is
\[
u(x,t)=\sum_{i=1}^N~-2k_i^2 {\rm sech} ^2(k_ix-4k_i^3t \pm \Delta_i)~,
\]
and the phase shifts are well-known
\cite{Drazin,Lamb,Wadati72} to be
\begin{equation}\label{phase}
e^{2 \Delta_n} = \prod_{m=1(m \neq n)}^{N}
\left| \frac{k_n-k_m}{k_n+k_m} \right|^{{\rm sgn}(n-m)}.
\end{equation}
For our case of $N=2$, one gets $\Delta_1=\delta$, $\Delta_2=-\delta$, where
\begin{equation}\label{phase11}
\delta \equiv \frac{1}{2} \ln \bigg [ \frac {k_2+k_1}{k_2-k_1} \bigg ]~~~.
\end{equation}
The general condition resulting from uniform
motion of the overall center of mass of a system is
\begin{equation}\label{cmcond}
\sum_{i=1}^N \frac {M_i \Delta_i}{k_i} = 0 .
\end{equation}
Our results for $\Delta_1$ and $\Delta_2$ are consistent with
Eq. (\ref{cmcond}) since  $M_1=-4k_1$ and $M_2=-4k_2$.
Note that since $k_2>k_1$, the Wronskian $W(\chi_1,\phi_2)$ produces a
solution $[S^0][S^0]$ with two singularities at large values of time.
The case of $[C^0][S^0]$ scattering is shown in Fig. 10. We have
checked from the graphs that the phase shifts are the same as in
Eq. (\ref{phase11}). Indeed all four entries in Table 3 are found to
have the same
phase shifts. This result is very plausible, since as mentioned in
Sec. 2, $C$-type and $S$-type negatons
of any given order are related to each other via an unphysical choice of phase
$k \xi=i \pi /2$, but this does not affect the scattering phase shift
calculation.

Next consider the scattering of a negaton of order 0 with a negaton of
order 1. Here one has the eight possibilities shown in Table 4. As expected,
all situations have the same phase shifts. More specifically, if one
considers the scattering of a soliton $[C^0]$ with wave
number $k_1$ and a negaton
$[S]$ with wave number $k_2>k_1$, the Wronskian is
$W(\phi_1,\phi_2,\partial_{k_2} \phi_2)$ and the soliton
gets phase shifted by
$\Delta_1=2\delta$, where $\delta$ is given in Eq. (\ref{phase11}).
This corresponds to the special case
of $N=3$ and $k_3 \rightarrow k_2$ in the general formula
Eq. (\ref{phase}), as might be expected from our physical
picture of a negaton. In contrast, the negaton $[S]$ gets phase shifted by
$\Delta_2=-\delta$. This follows from the center of mass
condition Eq. (\ref{cmcond}) with masses
$M_1=-4k_1$ and $M_2=-8k_2$ which were computed before.

It is clear that we can extend the above discussion to
the scattering of two negatons of order 1. For two
$[C]$ negatons, the Wronskian is
$W(\phi_1, \partial_{k_1} \phi_1, \phi_2, \partial_{k_2} \phi_2)$
which can be expanded to give:
\[
W_{CC} = [\gamma_1\gamma_2 k_1k_2 +{1\over 2} \gamma_1 k_1
\sinh 2\theta_2+{1\over 2}\gamma_2 k_2 \sinh 2\theta_1] (k_1^2-k_2^2)^2
\]
\begin{equation}\label{5.3}
+ {1\over 4} (k_1^4+6k_1^2k_2^2+k_2^4) \sinh 2\theta_1 \sinh 2\theta_2-4k_1k_2
(k_2^2 \cosh^2\theta_1 \sinh^2\theta_2+k_1^2 \sinh^2\theta_1 \cosh^2\theta_2).
\end{equation}
The Wronskians for the $SS$, $CS$ and  $SC$
negaton scattering solutions can similarly be
obtained.

At this stage, we can state the general result for phase shifts which
follows from the center of mass condition. Consider the scattering
of any negaton of order $n_1$ [wave number $k_1$, mass $M_1=-4k_1(n_1+1)$]
with a negaton of order $n_2$ [wave number $k_2$, mass $M_2=-4k_2(n_2+1)$].
Negaton $1$ will undergo a phase shift $\Delta_1=(n_2+1)\delta$
whereas negaton
$2$ will have a phase shift $\Delta_2=-(n_1+1)\delta$, with $\delta$ given by
Eq. (\ref{phase11}).
\sss

{\bf Positon-Negaton Scattering:} Here, we consider the
scattering of a $[\tilde C]$ positon with  negatons of different types.
The simplest situation is positon-soliton
$[\tilde C][C^0]$ scattering. The Wronskian is
$W(\tilde \phi, \partial_{\tilde k} \tilde \phi, \phi)$.
Matveev \cite{Matveev92} has shown
that for this case, the soliton has zero phase shift. In our approach,
the unchanged phase of
the soliton can be immediately and simply
understood from the center of mass condition
Eq. (\ref{cmcond}) and the fact that the positon $[\tilde C]$
has zero mass. The positon phase found by Matveev \cite{Matveev92} is
\begin{equation}\label{Deltap}
\Delta_p= \frac {1}{2} \tan^{-1} [2k \tilde k/(k^2-\tilde k^2)].
\end{equation}

Proceeding in the same way, the $[\tilde C][C]$
Wronskian $W(\tilde \phi,\partial_{\tilde k} \tilde \phi,
\phi, \partial_k \phi)$ is given by
\[
 W= (k^2+\tilde k^2)^2[k \tilde k \gamma \tilde \gamma
+ \frac {1}{2} k \gamma \sin 2\tilde \theta-\frac{1}{2}\tilde k
\tilde \gamma \sinh 2 \theta] +k \tilde k (k^2+\tilde k^2)
[\cosh 2\theta + \cos 2 \tilde \theta]
\]
\begin{equation}\label{5.5}
-\frac{1}{4}(k^4-6k^2 \tilde k^2+\tilde k^4)\sinh 2
\theta \sin2 \tilde \theta
+k \tilde k (k^2-\tilde k^2)[1+\cosh2 \theta \cos2 \tilde \theta].
\end{equation}

The scattering process is shown in
Fig. 11. Recall that the positon $[\tilde C]$ has zero mass, whereas the
negaton $[C]$ has mass $-8k$, twice the mass $-4k$ of a soliton $[C^0]$.
Therefore, our center of mass considerations predict that
the negaton will have zero phase shift and the
positon will have a phase shift $2 \Delta_p$, where $\Delta_p$ is given by
Eq. (\ref{Deltap}). We have confirmed this result by careful examination of
Fig. 11. Indeed, we can now state the general result
for a positon $[\tilde C]$ scattering
with any negaton of order $n$ [mass $-4k(n+1)$]. This scattering process
gives zero phase shift for the negaton
and $(n+1) \Delta_p$ for the positon.
\sss

\centerline{\bf 6. Singular Solutions of mKdV Equation}\sss

Recently, it has been shown \cite{Stahlhofen92} that the concept of
negatons and positons can
also be extended to the modified KdV equation:
\begin{equation}\label{6.1}
v_t - 6 v^{2}v_x +v_{xxx} = 0.
\end{equation}
If the KdV equation solutions $u(x,t)$ are given by Eqs. (\ref{uxt}) and
(\ref{Wn}),
then the corresponding
solution $v(x,t)$ of the mKdV equation
is given by \cite{Gesztesy89}
\begin{equation}
v(x,t) = \pm {\partial\over \partial x}
\ln \bigg [{W^*\over W} \bigg ],
\end{equation}
where
\begin{equation}\label{6.4}
W\equiv W(\phi, \partial_{k}\phi,...)\ ; ~~
W^* \equiv W(\phi, \partial_{k}\phi,...,1)~~.
\end{equation}
Thus, given a Wronskian $W$ (and hence $u$) of the
KdV equation, one can immediately obtain
the corresponding solution $v(x,t)$ of the mKdV equation by further computing
the Wronskian $W^*$. We would like to point out that
it is in fact
unnecessary to calculate $W^*$ since it can be shown to be related to $W$.
For example,
for the negaton solutions of order $n$ as given by $[C^{n}]$ and $[S^{n}]$
one can show that
\begin{equation}\label{6.5}
W^* [C^{n}] = k^{n+1} W[S^{n}], ~~W^* [S^{n}] = k^{n+1} W[C^{n}].
\end{equation}
Hence the negaton solutions of order $n$
of the mKdV equation are simply given by
\begin{equation}\label{6.6}
v = \pm {\partial\over \partial x}
\ln \bigg [ {W[S^{n}]\over W[C^{n}]}\bigg ].
\end{equation}
The singularities of $v$ come from the zeros of
$W[S^{n}]$ and $W[C^{n}]$, which
have been discussed previously, see Eq. (\ref{poles}). Therefore, the mKdV
negaton of order $n$ has $(n+1)$ singularities. In particular, there are no
nonsingular negaton solutions of the mKdV equation!

Using the $[C]$ and $[S]$ negaton Wronskians as given by
Eqs. (\ref{wc})
and (\ref{ws}), we find that the negaton solutions of order 1
of the mKdV equation
are given by
\begin{equation}\label{6.7}
v(x,t)=\pm \frac{4k(\sinh 2\theta-2k
\gamma \cosh 2\theta)} {(\sinh^{2} 2\theta -4k^2 \gamma^{2})}.
\end{equation}
Note that unlike the $[C^{n}]$ and $[S^{n}]$ negatons, the
corresponding negatons of the
mKdV equation differ from each other simply by a sign. The case of the
$n=1$ negaton is plotted in Fig. 12. We would like to remark
here that contrary to the claim of Stahlhofen \cite{Stahlhofen92}, the
negaton (or the corresponding positon) solutions (\ref{6.7})
of the mKdV equation do not lead to any new types of solutions
of the KdV equation
via the Miura transformation
\begin{equation}\label{6.8}
u_{1,2} = v^2 \pm v'~,
\end{equation}
but as expected, they simply give back the negaton solutions given by
Eqs. (\ref{ucc}) and (\ref{uss}).

 From the negaton-negaton scattering solutions of the KdV equation for wave
numbers $k_1$ and $k_2$ one
finds that
\begin{equation}\label{6.9}
W^*_{CC}= k_1^2k_2^2 W_{SS}, W^*_{SS}
= k_1^2k_2^2 W_{CC}, W^*_{CS} =
 k_1^2k_2^2 W_{SC}, W^*_{SC} =  k_1^2k_2^2 W_{CS},
\end{equation}
so that the negaton-negaton scattering solutions of the
mKdV equation are given by
\begin{equation}\label{6.10}
 v =  \pm {\partial\over \partial x}
\ln \bigg [ {W_{SS}\over W_{CC}}\bigg ]~,~
\pm {\partial\over \partial x} \ln \bigg [ {W_{SC}\over W_{CS}}\bigg ].
\end{equation}

Similarly, for the case of positon-negaton scattering, we
have the relations
\[
W^*_{\tilde C C} = \tilde k^2 k^2 W_{\tilde C S}
(\tilde \theta \rightarrow \tilde \theta +\pi/2),
\]
\begin{equation}\label{6.12}
W^*_{\tilde C S} =  \tilde k^2 k^2 W_{\tilde C C}
(\tilde \theta \rightarrow \tilde \theta +\pi/2)~,
\end{equation}
and hence the positon-negaton scattering solutions of the mKdV equation are
given by
\begin{equation}\label{6.13}
 v =  \pm {\partial\over \partial x}
\ln \bigg [ {W_{\tilde C S} (\tilde \theta\rightarrow
\tilde \theta +\pi/2)\over W_{\tilde C C}}\bigg ]~,
\end{equation}
\begin{equation}\label{6.14}
 v =  \pm {\partial\over \partial x}
\ln \bigg [ {W_{\tilde CC} (\tilde \theta \rightarrow
\tilde \theta +\pi/2)\over W_{\tilde C S}}\bigg ].
\end{equation}

Finally, for the positon-positon scattering case corresponding to wave
numbers $\tilde k_1$ and $\tilde k_2$, we have the relation
\begin{equation}\label{6.15}
 W^*_{\tilde C \tilde C}
= \tilde k_1^2 \tilde k_2^2 W_{\tilde C \tilde C} (\tilde
\theta_{1,2}\rightarrow
\tilde \theta_{1,2}+\pi/2)~,
\end{equation}
so that the positon-positon scattering solution of the mKdV equation is given
by
\begin{equation}\label{6.16}
 v  =  \pm {\partial\over \partial x}
\ln \bigg [
{W_{\tilde C \tilde C}(\tilde \theta_{1,2}\rightarrow \tilde
\theta_{1,2}+\pi/2)\over W_{\tilde C \tilde C}}\bigg]~.
\end{equation}
\sss

\centerline{\bf 7. Conclusions and Open Problems}\sss

In this paper we have discussed in some detail the properties of
negaton and positon solutions of the
KdV equation. Negaton solutions are quite different and at least
as interesting as the previously studied positon
solutions \cite{Matveev92,Matveev94,Stahlhofen95}. In particular, there
are two distinct types
of negaton solutions, whereas there is just one type of positon
solutions. We have also shown that using the KdV results one
can easily obtain the corresponding solutions of the mKdV equation.
There are several open problems which
are worth investigating. For example, in this paper we have chosen
a zero background potential. It would be worthwhile to see
if new phenomena arise with non-zero backgrounds. For the case of a
constant background potential,
there are well known nonsingular soliton
solutions of the KdV and mKdV equations which tend to non-zero values
as $x \rightarrow \pm \infty$, and singular solutions of the type
described in this paper can be readily constructed. It is expected that
these solutions will have some different properties since it will be
possible to have negatons moving to the left, in contrast to the
situation discussed in this paper. This should provide interesting
modifications of the scattering solutions of
positons and negatons. Another open problem
is to extend this analysis to the Dirac equation
\cite{Bordag77}.
We hope to address
these questions in the near future.
\sss

This work was supported in part by the U. S. Department of Energy under grant
DE-FG02-84ER40173. One of us (U.S.) would like to acknowledge the warm
hospitality of the Institute of Physics, Bhubaneswar, where this work was
begun and partially carried out.

\newpage

\noindent{\Large \bf Figure Captions}
\sss

\noindent {\bf Fig. 1:} The shape and motion of the $[C]$
negaton as given by
Eq. (\ref{ucc}) for $k=0.5$ and $\xi(k)=0$.
\sss

\noindent {\bf Fig. 2:} The (a) position $x_P$ and (b) velocity $v$ of
the singularity of the $[C]$ negaton of Fig. 1
as a function of time.
\sss

\noindent {\bf Fig. 3:} The positions  $x_Z$ of the two zeros of
the $[C]$ negaton of Fig. 1 as a function of time.
\sss

\noindent {\bf Fig. 4:} The shape and motion of
the $[S]$ negaton as given by
Eq. (\ref{uss}) for $k=0.5$ and $\xi(k)=0$.
\sss

\noindent {\bf Fig. 5:} The (a) position $x_P$ and (b) velocity $v$
of the singularity of the $[S]$ negaton of Fig. 4
as a function of time.
\sss

\noindent {\bf Fig. 6:} The positions of the two zeros  $x_Z$ of
the $[S]$ negaton of Fig. 4
as a function of time.
\sss

\noindent {\bf Fig. 7:} The shape and motion of
the $[S^4]$ negaton for $k=0.5$ and $\xi(k)=0$. Note that although
only 4 zeros are manifestly visible for the scales used in the figure,
there are indeed a total of 8 zeros corresponding to the
formula $2n$ discussed in Sec. 3.
\sss

\noindent {\bf Fig. 8:} The (a) position $x_P$ and (b) velocity $v$
of the three singularities of the $[S^4]$ negaton of Fig. 7
as a function of time.
\sss

\noindent {\bf Fig. 9:} The position  $x_P$ of
the singularity of the $[\tilde C]$ positon with $\tilde k=0.5$
as a function of time. Also shown is the line $x_P=-12 \tilde k^2 t$, with
a slope $-12 \tilde k^2$
which corresponds to the average speed of the singularity. The straight
sections have a slope $-8 \tilde k^2$.
\sss

\noindent {\bf Fig. 10:} Scattering of a soliton $[C^0]$ with wave number
$k_1=0.5$ and a $[S^0]$ negaton with wave number $k_2=1.0$.
\sss

\noindent {\bf Fig. 11:} Positon-negaton scattering. The
positon $[\tilde C]$ has a wave number $\tilde k = 1.0$ and the
negaton $[C]$ has a wave number $k=0.5$.
\sss

\noindent {\bf Fig. 12:} The shape and motion of
the order 1 negaton of the mKdV equation as given by
Eq. (\ref{6.7}) for $k=0.5$ and $\xi(k)=0$.

\newpage

\noindent{\Large \bf Table Captions}
\sss

\noindent {\bf Table 1:} Various characteristics of $[C^{n}]$
and $[S^{n}]$ negatons for $n=0,1,2,3$.
\sss

\noindent {\bf Table 2:} Various characteristics of $[\tilde C^n]$
positons for $n=0,1,2,3$.
\sss

\noindent {\bf Table 3:} Various possibilities for the scattering
of two negatons of order 0 and wave numbers $k_1$
and $k_2>k_1$. The quantities $\phi_i$ and $\chi_i$ are
defined in Eq. (\ref{phi}).
\sss

\noindent {\bf Table 4:} Scattering of a negaton of order 0 with a
negaton of order 1.
\sss

\newpage

\begin{table}
\begin{tabular}{||c|c|c|c|c|c||} \hline
Order & Negaton & {} & Dominant term & Poles & Zeros \\
of & type & Wronskian & in Wronskian & of & of \\
Wronskian & {} & $W(x,t)$ & at $x\rightarrow \pm \infty$ &
$u(x,t)$ & $u(x,t)$ \\ \hline
1 & $[C^0]$ & $\cosh\theta$ & $\cosh\theta$ & 0 & 0 \\
1 & $[S^0]$ & $\sinh\theta$ & $\sinh\theta$ & 1 & 0 \\
2 & $[C]$ & $k\gamma+\cosh\theta\sinh\theta$ &
$\cosh\theta\sinh\theta$ & 1 & 2 \\
2 & $[S]$ & $-k\gamma+\sinh\theta\cosh\theta$ &
$\sinh\theta\cosh\theta$ & 1 & 2 \\
3 & $[C^2]$ & Eq. (\ref{wc3}) &
$\cosh^2\theta\sinh\theta$ & 1 & 4 \\
3 & $[S^2]$ & Eq. (\ref{ws3}) &
$\sinh^2\theta\cosh\theta$ & 2 & 4 \\
4 & $[C^3]$ & Eq. (\ref{wc4}) &
$\cosh^2\theta\sinh^2\theta$ & 2 & 6 \\
4 & $[S^3]$ & Eq. (\ref{ws4}) &
$\sinh^2\theta\cosh^2\theta$ & 2 & 6 \\ \hline
\end{tabular}
\end{table}
\begin{table}
\begin{tabular}{||c|c|c|c||} \hline
Order & Positon & {} & Poles \\
of & type & Wronskian & of \\
Wronskian & {} & $W(x,t)$ & $u(x,t)$ \\ \hline
1 & $[\tilde C^0]$ & $\cos \tilde \theta$ & $\infty$ \\
2 & $[\tilde C]$ & $-\tilde k\tilde \gamma-\cos \tilde
\theta \sin \tilde \theta$
& 1 \\
3 & $[\tilde C^2]$ & Eq. (\ref{wtc3}) & $\infty$ \\
4 & $[\tilde C^3]$ & Eq. (\ref{wtc4}) & 2 \\ \hline
\end{tabular}
\end{table}
\begin{table}
\begin{tabular}{||c|c|c||} \hline
State & Number & {} \\
at & of & Wronskian \\
$t \rightarrow \infty$ & poles & $W(x,t)$ \\ \hline
$[C^0][C^0]$  & 0 & $W(\phi_1, \chi_2)$ \\
$[S^0][S^0]$  & 2 & $W(\chi_1, \phi_2)$ \\
$[C^0][S^0]$  & 1 & $W(\phi_1, \phi_2)$ \\
$[S^0][C^0]$  & 0 & $W(\chi_1, \chi_2)$ \\ \hline
\end{tabular}
\end{table}
\begin{table}
\begin{tabular}{||c|c|c||} \hline
State & Number & {} \\
at & of & Wronskian \\
$t \rightarrow \infty$ & poles & $W(x,t)$ \\ \hline
$[C][S^0]$  & 2 & $W(\phi_1,\partial_{k_1} \phi_1, \chi_2)$ \\
$[C^0][C]$  & 1 & $W(\phi_1, \chi_2, \partial_{k_2} \chi_2)$ \\
$[S][C^0]$  & 1 & $W(\chi_1,\partial_{k_1} \chi_1, \phi_2)$ \\
$[S^0][S]$  & 2 & $W(\chi_1, \phi_2, \partial_{k_2} \phi_2)$ \\
$[C][C^0]$  & 1 & $W(\phi_1,\partial_{k_1} \phi_1, \phi_2)$ \\
$[C^0][S]$  & 1 & $W(\phi_1, \phi_2, \partial_{k_2} \phi_2)$ \\
$[S][S^0]$  & 2 & $W(\chi_1,\partial_{k_1} \chi_1, \chi_2)$ \\
$[S^0][C]$  & 2 & $W(\chi_1, \chi_2, \partial_{k_2} \chi_2)$ \\ \hline
\end{tabular}
\end{table}

\newpage

\begin{thebibliography}{99}

\bibitem{Drazin} P. G. Drazin and R. S. Johnson, {\it Solitons},
Cambridge Univ. Press (1989).

\bibitem{Lamb} G. L. Lamb, {\it Elements of Soliton Theory}, John Wiley
and Sons (1980).

\bibitem{Eckhaus}  W. Eckhaus and A. Van Harten,
{\it The Inverse Scattering Transformation and the Theory of Solitons},
North-Holland (1981).

\bibitem{Kruskal74} M. D. Kruskal, Lect. Appl. Math. {\bf 15}, 61 (1974);
H. Airault, H. P. McKean and J. Moser, Comm. Pure Appl. Math. {\bf 30},
95 (1977);
M. J. Ablowitz and H. Cornille, Phys. Lett. {\bf A72},
277 (1979).

\bibitem{Matveev92} V. B. Matveev, Phys. Lett.  {\bf A166},
205 and 209 (1992); Phys. Lett. {\bf A168}, 463 (1992).

\bibitem{Matveev94} V. B. Matveev, J. Math. Phys. {\bf 35}, 2955 (1994).

\bibitem{Stahlhofen95} H. Maisch and A. A. Stahlhofen, to appear
in Physica Scripta (1995).

\bibitem{Stahlhofen92} A. A. Stahlhofen, Ann. Phys. {\bf 1}, 554 (1992).

\bibitem{Beutler93} R. Beutler, J. Math. Phys. {\bf 34}, 3098 (1993).

\bibitem{Beutler94} R. Beutler, A. Stahlhofen and V. B. Matveev,
Physica Scripta {\bf 50}, 9 (1994).

\bibitem{Matveev91} V. B. Matveev and M. A. Salle, {\it Darboux
Transformations and Solitons}, Springer-Verlag (1991).

\bibitem{Wadati72} M. Wadati and M. Toda, J. Phys. Soc. Japan {\bf 32},
1403 (1972).

\bibitem{Gesztesy89} F. Gesztesy, W. Schweiger and B. Simon,
preprint dedicated
to the memory of R. Hoegh-Krohn (1989).

\bibitem{Bordag77} L. Bordag, Vestnik Leningr.
Univers. ser. Math. Astr. and Mech.,
 No.2 (1977) (In Russian) ; D.B. Hinton, M. Klaus and J.K. Shaw, Proc. London
Math. Soc. {\bf 62}, 607 (1991) ; H. Behncke, Manus.
Math. {\bf 71}, 163 (1991) ;
B.A. Arbuzov et al., JETP Lett. {\bf 50}, 236 (1989).

\end{thebibliography}
\end{document}